\documentclass{osa-article}
\journal{osac}
\articletype{Research Article}
\usepackage{siunitx}
\usepackage{amsmath}
\usepackage{mathtools}
\usepackage{graphicx}

\usepackage{booktabs}
\usepackage{threeparttable}
\usepackage{array}
\usepackage{makecell}
\newif\ifmarkup \markupfalse
\newcommand{\rev}[1]{\ifmarkup{\color{red}#1}\else#1\fi}

\usepackage{lineno}
\ifmarkup\linenumbers\fi
\begin{document}

\title{Ultra-broadband quantum infrared spectroscopy extending to 20 {\textmu}m}

\author{Yu Mukai,\authormark{1,2,3} Ryo Okamoto ,\authormark{1,2} and Shigeki Takeuchi,\authormark{1,2,*}}

\address{\authormark{1}Department of Electrical, Electronic, and Digital Science and Engineering, Kyotodaigakukatsura, Nishikyo-ku, Kyoto University, Kyoto 615-8510, Japan\\
\authormark{2}Photonic Quantum Sensing Science and Engineering Center, Graduate School of Engineering, Kyoto University, Kyoto, Japan\\
\authormark{3}Japan Science and Technology Agency, PRESTO, Gobancho, Chiyoda-ku, Tokyo 102-0076, Japan\\}

\email{\authormark{*}takeuchi@kuee.kyoto-u.ac.jp} 

\begin{abstract*}
Quantum infrared spectroscopy (QIRS), which exploits quantum interference between photon-pair generation processes, enables infrared spectroscopy using only visible and near-infrared (VNIR) light sources and detectors. Extending the spectral coverage of QIRS into the long-wavelength infrared region up to 20 {\textmu}m is crucial for the identification and characterization of a broad range of materials. However, implementing QIRS beyond 10 {\textmu}m has been challenging because the nonlinear materials used to date do not provide suitable photon-pair sources in this spectral region. Here, we design a wavelength-tunable photon-pair source based on a gallium selenide crystal and develop a QIRS system covering a broad spectral range from 8.6 to 20 {\textmu}m. Using this system, we measure the transmittance spectrum of a polystyrene film and observe characteristic absorption bands beyond 10 {\textmu}m relevant to identifying polystyrene and to probing local chain conformation. These results demonstrate that QIRS is capable of covering the entire molecular fingerprint region.
\end{abstract*}

\section{Introduction}
Infrared (IR) sensing is a powerful tool for non-destructive material characterization, playing an indispensable role in a wide range of applications, including environmental monitoring, industrial inspection, and medical diagnostics. In particular, the so-called ``molecular fingerprint region'', extending to wavelengths up to 20 {\textmu}m, is of great importance for material identification by IR spectroscopy because it contains rich absorption features arising from molecular vibrational modes. However, in conventional approaches such as Fourier-transform infrared spectroscopy (FTIR) \cite{Jacquinot1960,Connes1966}, achieving higher sensitivity and greater device miniaturization in this spectral region remains challenging because of the reliance on thermal light sources and the thermal-noise-limited performance of infrared detectors.

Over the past few decades, photonic quantum technologies have advanced rapidly across various fields, including information processing \cite{O’Brien2009,Okamoto2009,Kiyohara2020,Park2025}, communications \cite{Bouwmeester1997,Duan2001,Gisin2007,Pirandola2020}, and sensing applications \cite{Nagata2007,Giovannetti2011,Ono2013,Lemos2014,Okano2015,Genovese2016}. Recently, photonic quantum sensing in the IR region using non-degenerate photon pairs has attracted significant attention as a novel approach to overcoming the technical limitations of conventional methods. Quantum infrared spectroscopy (QIRS), which exploits quantum interference between photon-pair generation processes, enables the evaluation of the IR properties of a sample through the detection of visible and near-infrared (VNIR) photons. This allows spectroscopic measurements without the need for IR light sources or detectors, instead taking advantage of high-performance VNIR light sources and silicon-based detectors. Furthermore, QIRS offers the distinct advantage of eliminating the influence of thermal background radiation \cite{Mukai2022}, which is a major noise source in conventional IR measurements. Inspired by the early proof-of-principle demonstration of QIRS in the mid-infrared (MIR) region\cite{Kalashnikov2016}, several studies have been reported that further advance this technique \cite{Paterova2018,Kutas2020,Lindner2020,Mukai2021,Lindner2021,Kaufmann2022,Tashima2024,Kurita2025}. 

Despite considerable efforts to broaden the spectral coverage of QIRS, previous demonstrations have remained largely limited to wavelengths below approximately 10 {\textmu}m, mainly owing to the lack of suitable photon-pair sources. As a result, QIRS has not yet been extended into the spectrally important 10--20 {\textmu}m region, which provides valuable vibrational information for characterizing functional groups and backbone conformations in a broad class of molecular and polymeric materials \cite{Nakaoki1991,Tashiro1980,Tashiro1992,Okada2016}. \rev{This is also the region in which blackbody radiation from the surroundings most severely limits the sensitivity of conventional measurements, and hence the region where photonic quantum sensing has the most to offer.} In earlier studies, nonlinear crystals such as LiNbO$_{3}$\cite{Kalashnikov2016,Paterova2018}, LiTaO$_{3}$ \cite{Tashima2024}, and KTiOPO$_4$ \cite{Kaufmann2022} were used for efficient photon-pair generation via the spontaneous parametric down-conversion (SPDC) process, but the generated infrared photons were restricted to below approximately 5 {\textmu}m because of the intrinsic infrared absorption by these crystals. Recently, silver gallium sulfide (AgGaS$_2$, AGS) crystals have enabled the development of VNIR and far-infrared (FIR\rev{; here 8--20 {\textmu}m, often also termed long-wave infrared}) photon-pair sources operating beyond 5 {\textmu}m \cite{Kumar2021}, as well as their application to QIRS \cite{Mukai2022,Paterova2022}, including our previous demonstration over the 8--10.5 {\textmu}m range \cite{Mukai2022}. However, because the transparency window of AGS remains insufficient for accessing longer wavelengths, further extension across the 10--20 {\textmu}m region has remained challenging. \rev{Very recently, SPDC with idler wavelengths beyond 20 {\textmu}m has been inferred from the signal spectra of AgGaS$_2$ and BaGa$_2$GeS$_6$ crystals \cite{Zatsepin2026a,Zatsepin2026b}. However, because no photon-counting detector is available at 10--20 {\textmu}m, where the room-temperature thermal background peaks, neither the emission nor the coherence of such long-wavelength idler photons has been verified, and QIRS beyond 10.5 {\textmu}m has not been realized.}

In this study, we realized a broadband QIRS system that covers wavelengths up to 20 {\textmu}m with an octave-spanning measurement bandwidth by employing a gallium selenide (GaSe) crystal. \rev{GaSe combines a transparency window extending to about 22 {\textmu}m with one of the largest nonlinear coefficients among infrared crystals, but it is a layered material available as (001)-cleaved, uncoated plates, so that phase matching has to be obtained by oblique incidence. Within this constraint, we designed a Type-II SPDC process pumped at 785 nm in a 100-{\textmu}m-thick plate, chosen so that the phase-matching bandwidth exceeds 250 cm$^{-1}$ at a single angle setting and the full 500--1160 cm$^{-1}$ range is covered with three settings, and we verified the design through the measured signal spectra and the absolute photon-pair generation rate. We then constructed a Michelson-type nonlinear quantum interferometer that returns the strongly divergent far-infrared idler, emitted at 10--24 times the angle of the signal, to the crystal, and observed, to our knowledge for the first time, nonlinear quantum interference extending to 20 {\textmu}m; in this band the interference itself constitutes direct evidence of the generation and coherence of the photon pairs. Using this system, we measured the infrared transmittance spectrum of a polystyrene film over the 8.6--20 {\textmu}m range by quantum Fourier-transform infrared spectroscopy (QFTIR) and observed absorption bands associated with phenyl-group vibrations that are characteristic of polystyrene and useful for distinguishing it from related polymers and for probing local chain conformations \cite{Nakaoki1991}. This work thus fills an important gap in infrared photonic quantum sensing. Moreover, the GaSe-based source is not tied to QFTIR and is expected to open the way to a variety of modalities of photonic quantum sensing.}

\section{Wavelength-tunable VNIR-FIR photon-pair generation using GaSe crystal} \label{section:source}
In this section, we present the design of a wavelength-tunable photon-pair source based on a GaSe crystal for broadband QIRS. By measuring the emission spectra of the signal photons in the VNIR region, we experimentally verify broadband tunability of the SPDC wavelengths via angle tuning and confirm good agreement between the measured phase-matched wavelengths and theoretical predictions.

The GaSe crystal employed in this study is a nonlinear material well suited for VNIR-FIR photon-pair generation beyond 10 {\textmu}m. Table \ref{tab:crystal_comparison} compares GaSe with representative nonlinear crystals that have previously been used as photon-pair sources for QIRS in terms of transparency range, \rev{effective nonlinear coefficient for the phase-matching geometry employed in each work,} and demonstrated QIRS spectral range. Compared with the nonlinear materials employed to date, GaSe exhibits both a large nonlinear coefficient and an exceptionally broad transparency range extending from the visible to the FIR region. Although its infrared transparency range is often regarded as being limited to around 18 {\textmu}m \cite{Fernelius1994}, it has been reported that the absorption coefficient remains below 2.5 cm$^{-1}$ up to approximately 22 {\textmu}m \cite{Allakhverdiev2003}. Therefore, if the crystal thickness is sufficiently smaller than the corresponding absorption length, VNIR-FIR photon-pair generation should remain feasible even at wavelengths approaching 20 {\textmu}m.

\begin{table}[t]
\centering
\caption{Comparison of nonlinear crystals used for QIRS photon-pair sources.}
\label{tab:crystal_comparison}
\begin{threeparttable}
\small\setlength{\tabcolsep}{5pt}
\renewcommand{\arraystretch}{1.2}
\begin{tabular}{lcccc}
\toprule
Crystal
& \makecell[c]{Transparency range \\ ({\textmu}m)}
& \rev{\makecell[c]{Phase matching}}
& \rev{\makecell[c]{$d_{\rm eff}$\tnote{f} \\ (pm/V)}}
& \makecell[c]{Demonstrated QIRS range \\ ({\textmu}m)} \\
\midrule
LiNbO$_3$
& \makecell[c]{0.33\tnote{a}  --5\tnote{b}}
& \rev{Type-I (e $\rightarrow$ oo), $\varphi = 30^\circ$}
& \rev{$\sim 5$\tnote{g}}
& \makecell[r]{1.9--5.2\cite{Arahata2022}}\\
\addlinespace[3pt]\cmidrule(lr){1-5}\addlinespace[3pt]
AgGaS$_2$
& \makecell[c]{ 0.50--13.2\tnote{c}}
& \rev{\makecell[c]{Type-I (e $\rightarrow$ oo), $\varphi = 45^\circ$ \\ Type-II (e $\rightarrow$ oe), $\varphi = 0^\circ$}}
& \rev{\makecell[c]{$\sim 9$\tnote{h} \\ $\sim 12$\tnote{h}}}
& \makecell[r]{8.0--10.5 \cite{Mukai2022}\\7.32--8.37 \cite{Paterova2022}}\\
\addlinespace[3pt]\cmidrule(lr){1-5}\addlinespace[3pt]
GaSe
& \makecell[c]{\begin{tabular}{@{}r@{--}l@{}}
0.65 & 18 \tnote{d}\\
  & 22.2 \tnote{e}
\end{tabular}}
& \rev{Type-II (e $\rightarrow$ oe), $\varphi = 0^\circ$}
& \rev{$\sim 50$\tnote{i}}
& \makecell[r]{8.6--20 (this work)}\\
\bottomrule
\end{tabular}

\begin{tablenotes}[flushleft]
\footnotesize
\item[]
\tnote{a}Cut-off wavelength at the absorption coefficient $\alpha \simeq 2\,\mathrm{cm}^{-1}$ for congruent LiNbO$_3$ (LN) \cite{Redfield1974}. \tnote{b}Cut-off wavelength at $\alpha \simeq 1\,\mathrm{cm}^{-1}$ for extraordinary wave in congruent LN \cite{Myers1996}.
\tnote{c}Transparency range defined as $\alpha < 3\,\mathrm{cm}^{-1}$ \cite{Bhar1972}.
\tnote{d}Transparency range defined as $\alpha < 1\,\mathrm{cm}^{-1}$ \cite{Fernelius1994},
\tnote{e}$\alpha < 2.5\,\mathrm{cm}^{-1}$ \cite{Allakhverdiev2003}.
\rev{\tnote{f}Effective nonlinear coefficient for the phase-matching geometry used in each work, evaluated from the tensor elements in notes g--i.
\tnote{g}$d_{31}=-4.6$ pm/V and $d_{22}=2.5$ pm/V (SHG at 1.06 {\textmu}m \cite{Miller1971}); $d_{\rm eff}=d_{31}\sin\theta_{\rm p}-d_{22}\cos\theta_{\rm p}\sin 3\varphi$ for $\theta_{\rm p} \sim 45^\circ$--$63^\circ$ in Ref.~\cite{Arahata2022}.
\tnote{h}$d_{36}=12.4$ pm/V (SFG at 0.778 {\textmu}m with input wavelengths of 10.2 and 0.842 {\textmu}m \cite{Zondy1997}); $d_{\rm eff}=d_{36}\sin\theta_{\rm p}\sin 2\varphi$ for Type-I ($\theta_{\rm p} \sim 44^\circ$--$50^\circ$ in Ref.~\cite{Mukai2022}) and $d_{\rm eff}=d_{36}\sin 2\theta_{\rm p}\cos 2\varphi$ for Type-II ($\theta_{\rm p} \sim 49^\circ$--$54^\circ$ in Ref.~\cite{Paterova2022}).
\tnote{i}$d_{22}=54$ pm/V \cite{Dmitriev1999}; $d_{\rm eff}=d_{22}\cos^{2}\theta_{\rm p}$ for $\theta_{\rm p}=13.7^\circ$--$18.1^\circ$.}
\end{tablenotes}
\end{threeparttable}
\end{table}

\begin{figure}[t]
\centering\includegraphics[width=13cm]{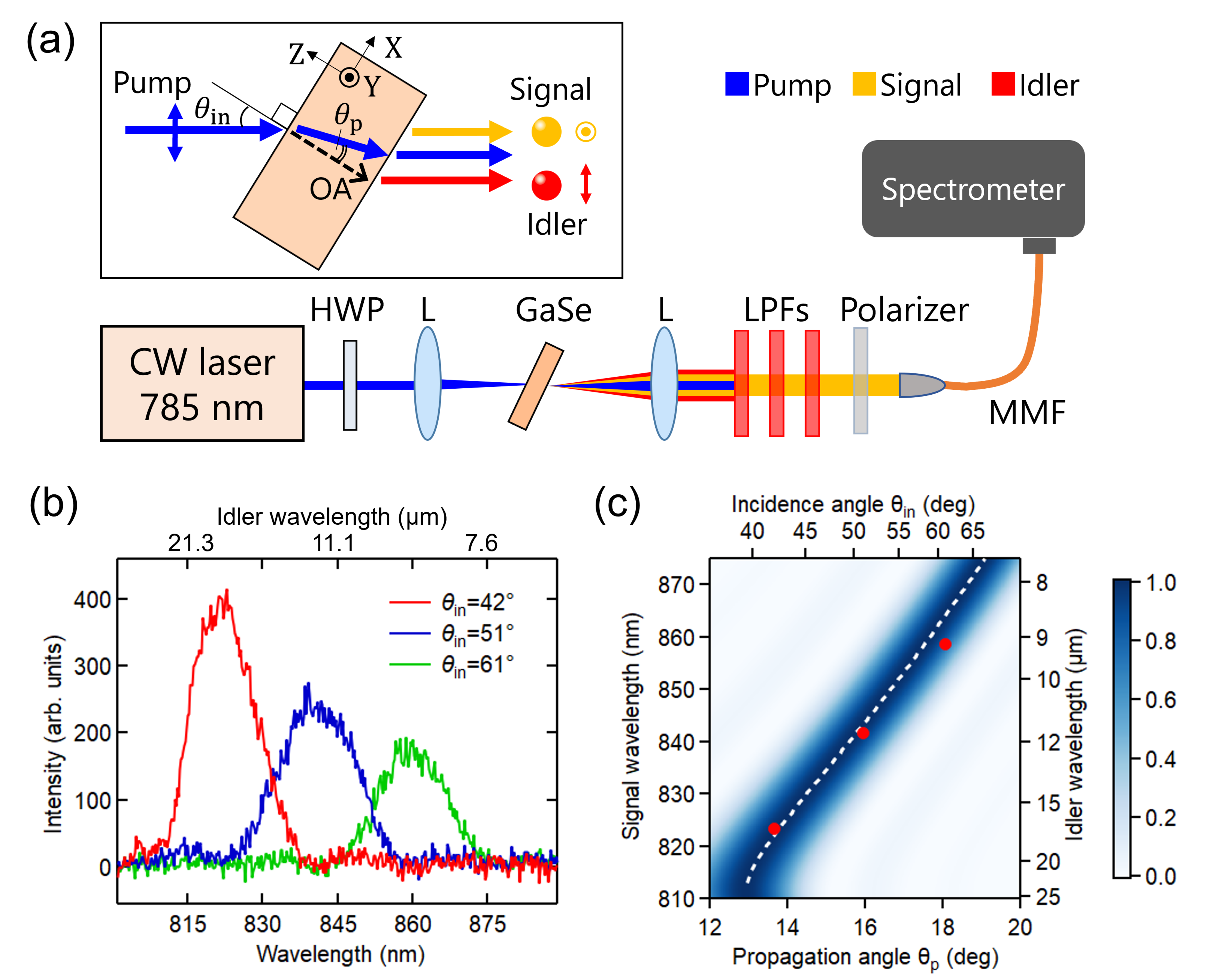}
\caption{ (a) Schematic of the experimental setup for measuring the SPDC emission spectra in VNIR region. HWP: half wave plate, L: lens ($f$ = 200 mm), LPF: long-pass filter (edge wavelength 800 nm), MMF: multi-mode fiber. The inset shows a conceptual diagram of photon-pair generation via Type-II SPDC in a GaSe crystal. The solid black arrows represent the crystallographic axes of the GaSe crystal. The dashed black arrow indicates the optic axis (OA). The pump and idler photons are polarized parallel to the plane of incidence (extraordinary rays), while the signal photon is polarized perpendicular to the plane (ordinary ray). $\theta_{\rm{in}}$ and $\theta_{\rm{p}}$ are the incidence and propagation angles of the pump light. (b) Emission spectra of signal photons generated via SPDC in the GaSe crystal. The upper horizontal axis shows the corresponding idler wavelength. \rev{(c) Calculated phase-matching function $\mathrm{sinc}^{2}(\varDelta k L/2)$ of the collinear SPDC process (color map) as a function of the propagation angle $\theta_{\rm{p}}$ and the signal wavelength; the upper and right axes show the corresponding incidence angle $\theta_{\rm{in}}$ and idler wavelength. The white dashed line indicates the perfectly phase-matched wavelength ($\varDelta k = 0$) and the red circles the spectral centroids of the measured spectra in (b). The map extends to an idler wavelength of 25 {\textmu}m; the Sellmeier equations of Ref.~\cite{Vodopyanov1995} are specified for 0.65--18 {\textmu}m, and the calculation beyond 18 {\textmu}m is an extrapolation.}}
\label{SPDCEvaluation}
\end{figure}

As schematically shown in the inset of Fig. \ref{SPDCEvaluation}(a), we utilize a collinear Type-II (e $\rightarrow$ oe) SPDC process, where the pump and idler are extraordinary rays and the signal is an ordinary ray. The GaSe crystal is cut parallel to the (001) plane, and birefringent phase matching is achieved through oblique incidence of the pump beam with respect to the [001] direction (the crystal Z-axis). The crystal X-axis is defined here as being parallel to the crystallographic [100] direction, and the experimental geometry is arranged such that the plane of incidence lies within the XZ-plane. The energy conservation and the phase-matching condition for the SPDC process are given by
\begin{equation}\label{EngCon}
 \hbar\omega_{\rm{p}}= \hbar\omega_{\rm{s}}+ \hbar\omega_{\rm{i}},
\end{equation}
\begin{equation}\label{PhMatch}
n_{\rm{e}}\left(k_{\rm{p}},\theta_{\rm{p}}\right)k_{\rm{p}}=n_{\rm{o}}\left(k_{\rm{s}}\right)k_{\rm{s}}+n_{\rm{e}}\left(k_{\rm{i}},\theta_{\rm{p}}\right)k_{\rm{i}},
\end{equation}
where $\hbar$ is Dirac's constant, $\omega$ is the angular frequency, $k$ is the wavenumber in a vacuum, and the subscripts p, s, and i denote the pump, signal and idler photons, respectively. $n_{\rm{o}}$ is the refractive index for the ordinary ray and $n_{\rm{e}}(k,\theta_{\rm{p}})$ is the effective refractive index for the extraordinary ray, which depends on the propagation angle $\theta_{\rm{p}}$. Here, $\theta_{\rm{p}}$ is the angle between the propagation direction of the light in the GaSe crystal and the optic axis (the crystal Z-axis), and is related to the incidence angle  $\theta_{\rm{in}}$ through Snell's law $\left(\sin\theta_{\rm{in}} = n_{\rm{e}}\left(k_{\rm{p}},\theta_{\rm{p}}\right) \sin\theta_{\rm{p}}\right)$. $n_{\rm{e}}(k,\theta_{\rm{p}})$ is related to $n_{\rm{o}}$ and the principal refractive index for the extraordinary ray $\bar{n}_{\rm{e}}$ as

\begin{equation}\label{ERayRef}
n_{\rm{e}}\left(k,\theta\right)=\left(\frac{\sin^{2}\theta}{\bar{n}^{2}_{\rm{e}}\left(k\right)}+\frac{\cos^{2}\theta}{n^{2}_{\rm{o}}\left(k\right)}\right)^{-\frac{1}{2}}.
\end{equation}


We experimentally measured the emission spectra of the VNIR signal photons generated in the GaSe crystal using the spectroscopic setup shown in Fig. \ref{SPDCEvaluation}(a). As the pump source, we used a single-frequency CW laser with a wavelength of 785 nm (output power 100 mW, linewidth < 1 MHz)\rev{; the pump power at the crystal, after fiber delivery and isolation, was 45 mW for the measurements described in this section}. The pump beam was focused onto the GaSe crystal (thickness 100 {\textmu}m, uncoated) by a lens ($f=200$ mm) \rev{to a $1/e^{2}$ diameter of 140 {\textmu}m}. The GaSe crystal is mounted on a rotation stage, allowing it to be rotated around the crystal Y-axis to adjust the incidence angle of the pump beam. The pump polarization was set to p-polarization to induce Type-II SPDC. The generated signal photons were collimated by a second lens ($f=200$ mm) and then passed through long-pass filters with a cut-on wavelength of 800 nm and a polarizer to remove the pump light and unwanted fluorescence from the GaSe crystal. The transmitted signal photons were subsequently coupled into a multimode fiber and sent to a VNIR spectrometer (SR500i + DU416A-LDC-DD, Andor, wavelength resolution 0.1 nm). \rev{The collection half-angle of the signal photons, 18.75 mrad, is determined by the 7.5-mm effective aperture of the fiber-coupling lens.}

Figure \ref{SPDCEvaluation}(b) shows the experimentally observed emission spectra of the signal photons. Measurements were performed at three different pump incidence angles: $\theta_{\rm{in}} = 42^\circ, 51^\circ,$ and $61^\circ$\rev{, corresponding to internal propagation angles of $\theta_{\rm{p}} = 13.7^\circ, 16.0^\circ,$ and $18.1^\circ$}. As expected from the angular dependence of the phase-matching condition in Eq. (2), the peak wavelength shifts significantly with the incidence angle of the pump light. Specifically, signal emission was observed between 815 nm and 870 nm, implying that the correlated idler photons were generated across a broad FIR range from 8 {\textmu}m to 20 {\textmu}m. \rev{Figure \ref{SPDCEvaluation}(c) shows the calculated phase-matching function $\mathrm{sinc}^{2}(\varDelta k L/2)$ of the collinear Type-II SPDC process under our experimental conditions as a function of the propagation angle $\theta_{\rm{p}}$ and the signal wavelength, where $\varDelta k$ is the phase mismatch corresponding to Eq.~(\ref{PhMatch}) and $L = 100$ {\textmu}m is the crystal length; the white dashed line indicates the perfectly phase-matched wavelength ($\varDelta k = 0$). Because the pump is a single-frequency CW laser, this function represents the joint spectral intensity of the collinearly emitted photon pairs.} The refractive indices of the GaSe crystal were determined using the Sellmeier equations provided in Ref.\cite{Vodopyanov1995}\rev{, which are specified for the 0.65--18 {\textmu}m range; the calculation at idler wavelengths beyond 18 {\textmu}m is therefore an extrapolation}. The three data points in Fig. \ref{SPDCEvaluation}(c) represent the spectral centroids of the measured signal spectra in Fig. \ref{SPDCEvaluation}(b), showing fair agreement with the theoretical predictions. These results confirm the generation of VNIR signal photons, indicating the successful production of highly nondegenerate photon pairs via SPDC in GaSe as designed. The correlated idler photons are expected to reach the FIR region at wavelengths up to 20 {\textmu}m.

\rev{To characterize the brightness of the source, we measured the signal count rate at the three incidence angles with the full collection aperture, coupling the multimode fiber to a silicon avalanche photodiode (SPCM-AQRH-14-FC, Excelitas Technologies) instead of the spectrometer. Table~\ref{tab:rates} compares the measured count rates $N$ with the theoretical expectation $R_{\rm th}$, the rate of signal photons emerging from the crystal within the collection cone calculated with the low-gain SPDC theory \cite{Ling2008}. The measured rates are corrected for the system detection efficiency $\eta$ of the signal path after the crystal, comprising the long-pass filters, the polarizer, the fiber coupling, and the detector efficiency ($\eta = 0.27$--$0.32$). The corrected rates $N/\eta$ agree with $R_{\rm th}$ to within 25\% for all three settings. Correcting further for the Fresnel losses at the crystal, the rate of signal photons generated within the collection cone amounts to $\approx 1 \times 10^{6}$ s$^{-1}$. With the phase-matching bandwidth exceeding 250 cm$^{-1}$ (biphoton correlation time $\approx 130$ fs), this corresponds to at most $\sim 10^{-7}$ photon pairs per correlation time even if all photons occupied a single spatial mode, confirming operation deep in the low-gain regime.}

\begin{table}[t]
\centering
\rev{
\begin{threeparttable}
\caption{Measured signal count rates and comparison with the theoretically expected pair-generation rate, corrected for the system detection efficiency $\eta$.}
\label{tab:rates}
\small\setlength{\tabcolsep}{3pt}
\begin{tabular}{lccccc}
\toprule
\makecell[c]{Incidence angle $\theta_{\rm in}$ \\ ($\lambda_{\rm s}$ / $\lambda_{\rm i}$)} & \makecell[c]{Measured rate \\ $N$ (s$^{-1}$)\tnote{a}} & \makecell[c]{System detection \\ efficiency $\eta$\tnote{b}} & \makecell[c]{Estimated rate \\ $N/\eta$ (s$^{-1}$)} & \makecell[c]{Theory \\ $R_{\rm th}$ (s$^{-1}$)\tnote{c}} & $N/(\eta R_{\rm th})$ \\
\midrule
$42^\circ$ (822 nm / 17.4 {\textmu}m) & $156\,530 \pm 380$ & $0.32 \pm 0.04$ & $(4.9 \pm 0.6)\times 10^{5}$ & $6.00 \times 10^{5}$ & $0.82 \pm 0.11$ \\
$51^\circ$ (843 nm / 11.3 {\textmu}m) & $169\,560 \pm 430$ & $0.30 \pm 0.04$ & $(5.6 \pm 0.7)\times 10^{5}$ & $7.39 \times 10^{5}$ & $0.75 \pm 0.10$ \\
$61^\circ$ (864 nm / 8.6 {\textmu}m)  & $125\,680 \pm 340$ & $0.27 \pm 0.04$ & $(4.6 \pm 0.6)\times 10^{5}$ & $5.99 \times 10^{5}$ & $0.77 \pm 0.11$ \\
\bottomrule
\end{tabular}
\begin{tablenotes}[flushleft]
\footnotesize
\item[]
\tnote{a}Single-pass configuration of Fig.~\ref{SPDCEvaluation}(a), pump power 45 mW at the crystal, collection half-angle 18.75 mrad. Mean $\pm$ standard deviation of 50 successive 1-s readings.
\tnote{b}System detection efficiency of the signal path after the crystal (see text); the uncertainty reflects the assumed polarizer and fiber-coupling efficiencies ($0.85 \pm 0.05$ and $0.70 \pm 0.05$).
\tnote{c}Calculated rate of signal photons emerging from the crystal within the collection angle, for Type-II phase matching with a pump $1/e^{2}$ diameter of 140 {\textmu}m, including the Fresnel losses at the uncoated crystal surfaces.
\end{tablenotes}
\end{threeparttable}}
\end{table}

\section{Ultra-broadband QIRS over the 8.6--20~{\textmu}m spectral range}\label{QIRS}
In this section, we construct a nonlinear quantum interferometer based on the GaSe-crystal photon-pair source and demonstrate broadband QIRS over the 8.6--20~{\textmu}m spectral range. We first characterize the quantum interferograms and the corresponding Fourier spectra obtained without a sample, confirming that the interferometer accesses infrared wavelengths extending to 20~{\textmu}m. We then apply the system to QFTIR measurements of a polystyrene film and evaluate its transmittance spectrum over the long-wavelength portion of the molecular fingerprint region.

\begin{figure}[h]
\centering\includegraphics[width=13cm]{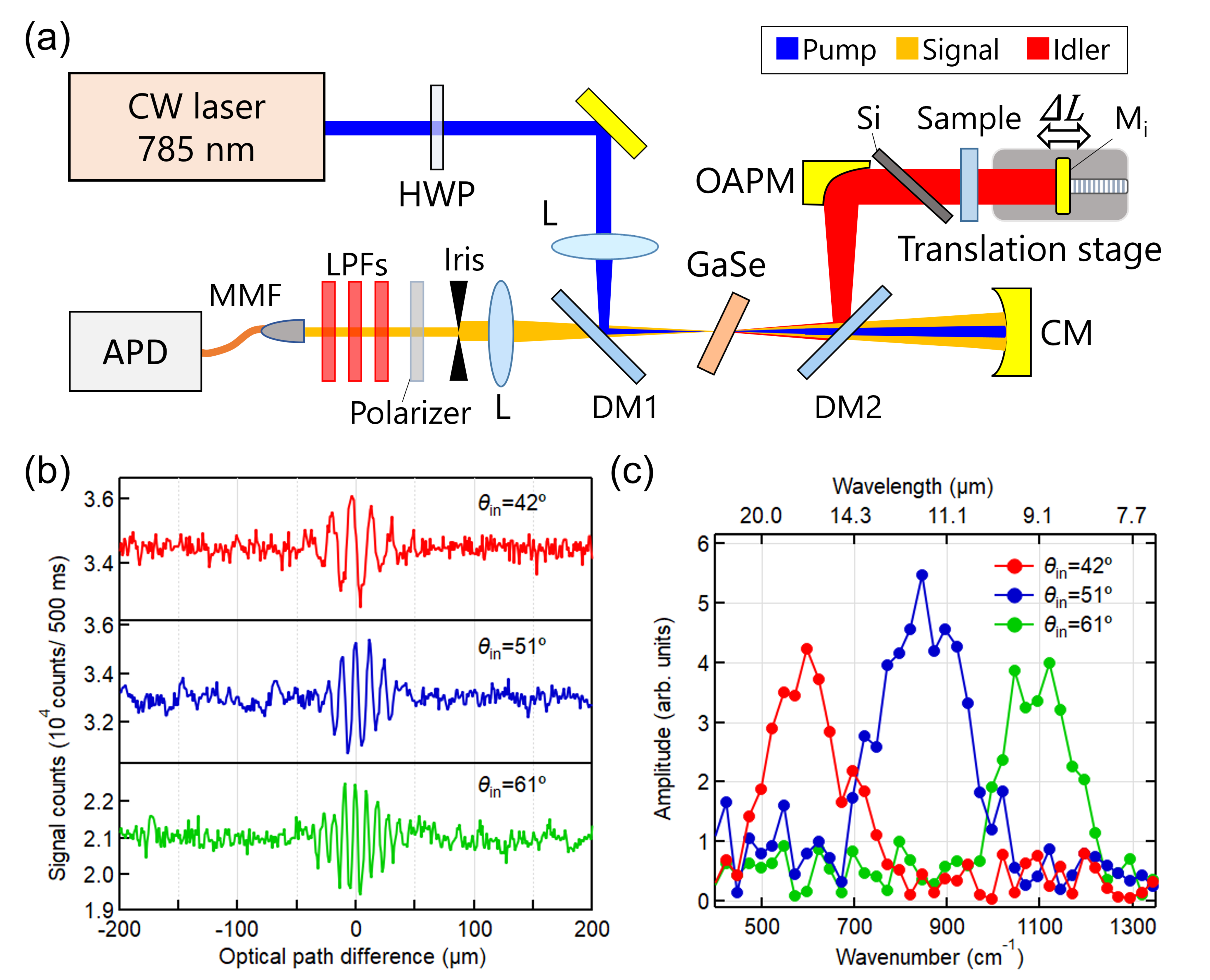}
\caption{(a) Schematic of the nonlinear quantum interferometer. HWP: half wave plate, L: lens ($f$ = 200 mm), OAPM: off-axis
parabolic mirror ($f$ = 152 mm), DM1: long pass dichroic mirror (edge wavelength 800 nm), DM2: short pass dichroic mirror (edge wavelength 2000 nm), CM: concave end mirror ($f$ = 150 mm), Mi: end mirror, LPF: long-pass filter (edge wavelength 800 nm), MMF: multi-mode fiber, APD: Si avalanche photodetector, Si: silicon Brewster window\rev{, Iris: 3-mm-diameter iris in the collimated section of the signal beam}. (b) Time-domain quantum interferograms and (c) Fourier amplitude spectra.}
\label{NLIntf}
\end{figure}

Figure \ref{NLIntf}(a) presents a schematic of the QIRS system using the photon-pair source described in the previous section. Our setup employs a Michelson-type nonlinear interferometer for QFTIR measurements, following the architecture reported in Refs.\cite{Mukai2022,Mukai2021}. The CW laser at 785 nm serves as the pump source\rev{, delivering 62 mW at the crystal,} focused into the GaSe crystal by a lens with a focal length of $f = 200$~mm \rev{under the same focusing conditions as in Sec.~\ref{section:source}}. The pump beam is horizontally polarized (corresponding to the extraordinary ray), which generates vertically polarized (ordinary) VNIR signal photons and horizontally polarized (extraordinary) FIR idler photons through the Type-II SPDC process. The idler photons are reflected by a dichroic mirror DM2 to be spatially separated from the pump and signal beams, then collimated by an off-axis parabolic mirror OAPM ($f = \rev{152}$~mm\rev{, 50.8 mm aperture}). To eliminate residual pump and signal light in the idler path, a 140 {\textmu}m-thick silicon (Si) wafer is positioned at the Brewster angle. The infrared light is then reflected by an end mirror $\rm{M_i}$ and refocused back into the GaSe crystal. Meanwhile, the pump and signal photons transmitted through DM2 are reflected by a concave mirror ($f = 150$~mm) and refocused into the crystal.

As a result of quantum interference between the SPDC processes occurring in the forward and backward passes, the photon-pair generation rate is modulated depending on the relative phase between the two processes. This relative phase is tuned by scanning the optical path length difference $\varDelta L$ between the infrared and VNIR arms, achieved by translating the end mirror M$_i$ using a high-precision translation stage (NFS100-50UPX, Sigma Koki). Finally, the signal photons passing through DM1 and the long-pass filters are coupled into a multi-mode fiber guided to a silicon avalanche photodiode APD (SPCM-AQRH-14FC, Excelitas Tech.), and the photon count is recorded with a photon counter
(SR400, Stanford Research Systems).  \rev{An iris of 3 mm diameter, placed in the collimated section of the signal beam, restricts detection to signal photons whose conjugate idler photons, emitted at angles $\lambda_{\rm i}/\lambda_{\rm s} \approx 10$--24 times larger owing to transverse-momentum anticorrelation, are collected and returned by the idler-arm optics.} The quantum interferogram is defined as the signal photon count rate $P_s(\varDelta L)$ as a function of the scan distance $\varDelta L$. In the QFTIR method, the spectral intensity of the idler photons contributing to the quantum interference is retrieved by performing a Fourier transform on the quantum interferogram $P_s(\varDelta L)$. The infrared transmittance spectrum is then obtained by comparing the resulting Fourier spectra measured with and without a sample in the idler path \cite{Mukai2021}.

First, using the constructed nonlinear quantum interferometer, we experimentally observed quantum interferograms in the long-wavelength FIR region. The signals were measured as reference data without a sample in the idler path. The measurement parameters included a step size of 1 {\textmu}m, a total scan distance $W$ of 400 {\textmu}m, and an integration time of 500 ms per point. Measurements were performed at three pump incidence angles: $\theta_{\rm{in}} = 42^\circ, 51^\circ,$ and $61^\circ$. As shown in Fig. \ref{NLIntf}(b), clear interference signals were successfully observed. At $\theta_{\rm{in}} = 61^\circ$, the average photon count rate was on the order of $2\times 10^{4}$ per 500 ms and the oscillation period was approximately 9 {\textmu}m. 
The interference visibility $V=(P_{\rm{max}}-P_{\rm{min}})/(P_{\rm{max}}+P_{\rm{min}})$ was approximately 7$\%$, where
$P_{\rm{max}}$ and $P_{\rm{min}}$ are the maximum and minimum photon count rates, respectively. \rev{Since the optical losses expected in the interferometer alone cannot account for the low visibility, additional factors such as momentum--spectrum entanglement of the broadband emission, uncompensated dispersion over the broad bandwidth, and spatial-mode mismatch between the forward and backward SPDC emission are also plausible.} As the incidence angle was varied, the oscillation period of the interferogram changed, reflecting the shift in the SPDC emission wavelength. At $\theta_{\rm{in}} = 42^\circ$, the oscillation period increased to 18 {\textmu}m.

Figure \ref{NLIntf}(c) shows the infrared spectra contributing to the quantum interference, retrieved by Fourier transforming the time-domain quantum interferograms. The wavenumber resolution of the Fourier spectra was $\varDelta  k= 1/W= \SI{25}{\per\centi\meter}$. For each condition, the emission bandwidth exceeded \SI{200}{\per\centi\meter}, with the infrared idler spectra distributed over 8 {\textmu}m to 20 {\textmu}m. These results clearly demonstrate that the GaSe crystal can serve as a photon-pair source covering the long-wavelength infrared region and can be effectively applied to QIRS. \rev{Since photon-counting detection of the idler is not available at these wavelengths, the nonlinear interference itself constitutes direct evidence of the generation and coherence of the photon pairs up to 20 {\textmu}m.}

\begin{figure}[ht]
\centering\includegraphics[width=13cm]{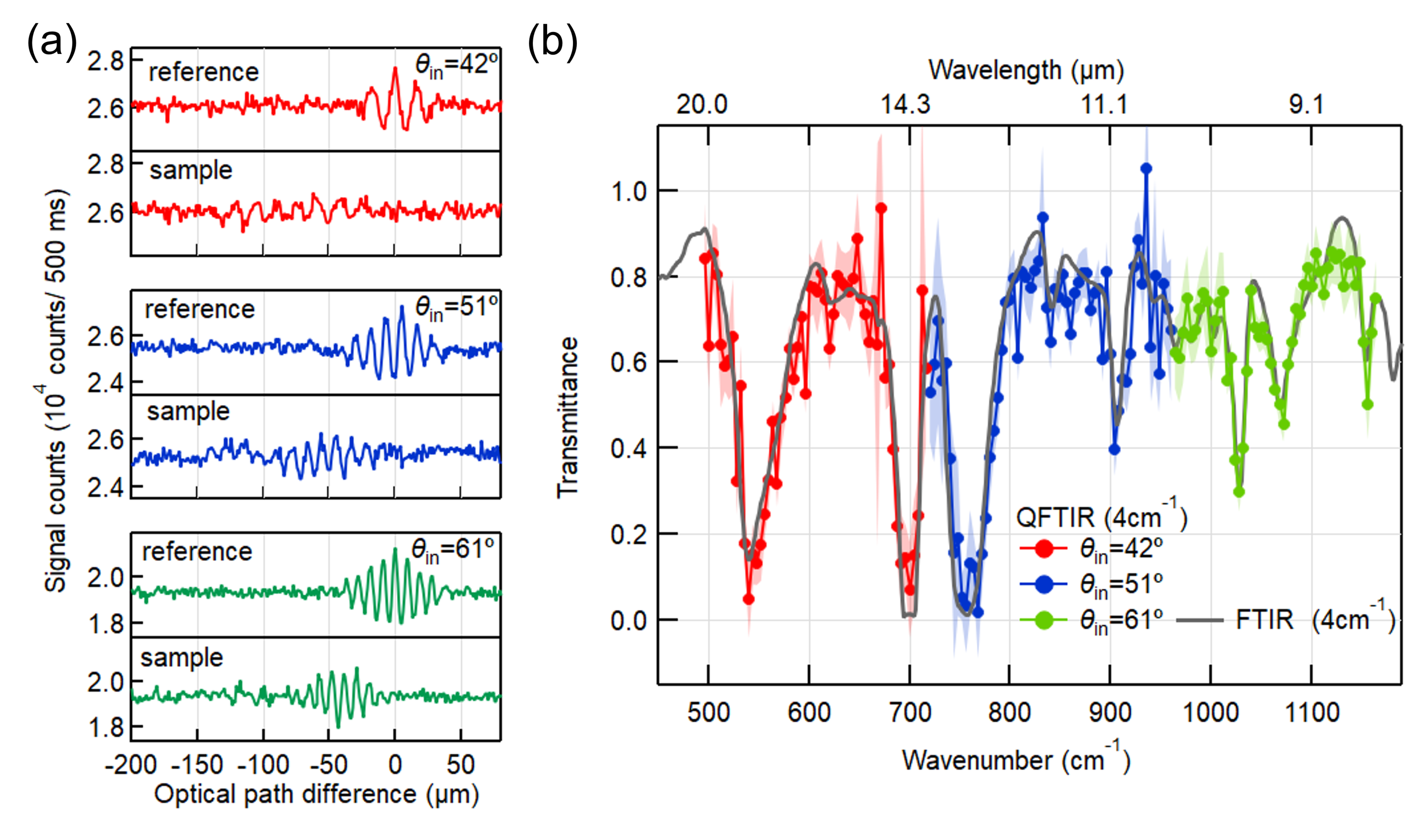}
\caption{(a) Time-domain quantum interferograms measured with and without the 30-{\textmu}m thick polystyrene film sample. (b) Transmittance spectrum of the sample measured by QFTIR method (\rev{red, blue, and green circles for $\theta_{\rm in} = 42^\circ$, $51^\circ$, and $61^\circ$; the shaded bands indicate the standard error over the 20 measurement sets}) and conventional FTIR spectrometer (gray line).}
\label{QFTIR}
\end{figure}

Next, we demonstrated QIRS in the long-wave FIR region using the developed system. A 30 {\textmu}m-thick polystyrene film was employed as a sample and its transmittance spectrum was measured via the QFTIR method. For the quantum interferogram measurements, the step size was set to 1 {\textmu}m, the integration time to 500 ms per point, and the total scan distance $W$ to 2500 {\textmu}m, resulting in a wavenumber resolution of $\varDelta  k\left(= 1/W\right)= \SI{4}{\per\centi\meter}$. To achieve broadband spectroscopy, measurements were performed at three different pump incidence angles. Figure \ref{QFTIR}(a) shows the time-domain quantum interferograms near the interference peak measured without and with the sample in the idler path, denoted as the reference and sample signals, respectively. \rev{The present measurements operate close to the shot-noise limit: the baseline noise of the reference interferograms, evaluated over 1900 fringe-free points ($\varDelta L = -2000$ to $-100$ {\textmu}m), exceeds the shot-noise level only by factors of 1.13, 1.25, and 1.17 for $\theta_{\rm in} = 42^\circ$, $51^\circ$, and $61^\circ$, respectively. We attribute this slight excess to pointing fluctuations of the pump laser, which shift the signal emission pattern on the 3-mm iris and thereby translate into fluctuations of the signal transmittance through the iris.} A pronounced modification of the interferogram waveform was observed when the sample was inserted, reflecting infrared absorption by the polystyrene film. The transmittance spectrum of the sample was calculated by taking the ratio of the Fourier spectrum of the sample and reference signal. Figure \ref{QFTIR}(b) shows the absolute transmittance spectra obtained by taking the absolute value of the averaged complex transmittance spectra measured using the QFTIR method. The averaging was performed over 20 \rev{reference--sample pairs of interferograms} for each of the three pump incidence angles, $\theta_{\rm in}=42^\circ$, $51^\circ$, and $61^\circ$\rev{; with approximately 21 min per interferogram, the 120 interferograms correspond to a total integration time of approximately 42 h. The shaded bands indicate the standard error over the 20 sets}. These spectra range from 1164 cm$^{-1}$ to 500 cm$^{-1}$, corresponding to wavelength of 8.6 {\textmu}m to 20 {\textmu}m.

We successfully observed the major infrared absorption bands characteristic of polystyrene over this long-wavelength spectral range. Notably, strong absorption peaks appeared at around 760 and 700 cm$^{-1}$, which are characteristic of out-of-plane C--H deformation modes of the monosubstituted benzene ring. A pronounced band was also observed around 540 cm$^{-1}$, which can be assigned to an out-of-plane deformation mode of the phenyl group and is sensitive to local chain conformation\cite{Nakaoki1991}. Additional peaks at approximately 1070 and 1030 cm$^{-1}$ are assigned to in-plane deformation modes of the phenyl C--H bonds, while a distinct band near 900 cm$^{-1}$ is attributed to an out-of-plane deformation mode.
The solid lines in the same figure represent the transmittance spectra measured using a conventional FTIR spectrometer (IRTracer-100, Shimadzu Corp.) \rev{with a resolution of 4 cm$^{-1}$ and 20 scans, acquired in approximately 1 min}. \rev{The QFTIR results agree with the conventional FTIR measurements within the statistical uncertainty over most of the spectral range.}

\rev{\section{Discussion} \label{section:discussion}
The broadband QFTIR demonstrated here establishes that GaSe operates as a photon-pair source in the long-wavelength infrared region. At present, however, the practical limitation of the system is the long acquisition time of the quantum interferograms (Sec.~\ref{QIRS}). Since the baseline noise is close to the shot-noise limit, the signal-to-noise ratio is determined by the number of detected signal photons and by the interference visibility; the acquisition time is thus governed mainly by the absolute number of photons effectively available for the nonlinear quantum interferometry. The source itself performs as designed: as shown in Sec.~\ref{section:source}, the measured count rates are in quantitative agreement with the theoretically expected photon-pair generation rate. In this work, a rather thin crystal of 100 {\textmu}m was employed so that a phase-matching bandwidth exceeding 250 cm$^{-1}$ is obtained even at a single angle setting, allowing the full 500--1160 cm$^{-1}$ range to be covered with only three settings. Since the phase-matching bandwidth scales inversely with the crystal length \cite{Ling2008}, a longer crystal would increase the spectral brightness ($\propto L^{2}$) but proportionally narrow the bandwidth, while visibility-degrading effects such as spatial walk-off and unbalanced dispersion also grow with the crystal length. A moderately thicker crystal of a few hundred micrometers may therefore offer a better compromise between pair flux and bandwidth.

A larger gain is available from the fraction of the generated photon pairs that actually contributes to the interference. The 3-mm iris, which matches the detected signal photons to the acceptance of the idler arm, discards a large fraction of the signal photons collected with the full aperture in Sec.~\ref{section:source}, because their conjugate idler photons, emitted at 10--24 times larger angles, are not returned to the crystal. Enlarging the acceptance of the idler arm, for example with larger-aperture or shorter-focal-length collection optics, would potentially allow a larger fraction of the generated photon pairs to contribute to the interference without compromising the visibility. Together with the optimization of the crystal thickness discussed above, this offers a realistic path toward shorter acquisition times.

The QFTIR scheme employed in the present demonstration has practical merits: the spectral resolution is set by the scan range and is not limited by the pump linewidth, so that broadband pulsed pump lasers can also be employed \cite{Kaur2024}; the setup remains compact with a single-pixel detector; and it extends naturally to hyperspectral imaging \cite{Placke2026}. From the viewpoint of the signal-to-noise ratio in spectroscopic measurements, on the other hand, spectral-domain acquisition has an intrinsic advantage over scan-based acquisition for shot-noise-limited measurements with VNIR detectors, as is the case here, because Fellgett's multiplex advantage vanishes, or even reverses, in this regime, as recently shown by Gattinger \textit{et al.} for scanless QFTIR \cite{Gattinger2026}. Since this advantage grows with the number of resolved spectral elements, it is particularly relevant for a broadband source such as ours. The acquisition scheme can thus be chosen according to the requirements of each application.}

\section{Conclusion} \label{section:conclusion}
In conclusion, we have successfully demonstrated QIRS over the broad FIR region ranging from 8.6 to 20 {\textmu}m, covering a significant portion of the molecular fingerprint region. By exploiting a SPDC process in a GaSe crystal, we developed a wavelength-tunable VNIR-FIR photon-pair source and constructed a Michelson-type nonlinear quantum interferometer. \rev{This setup enables the acquisition of quantum interferograms from which the infrared transmittance spectrum can be retrieved.} The performance of our system was validated through QFTIR measurements of a polystyrene film, for which the measured transmittance spectra accurately reproduced the characteristic absorption features of the sample and showed quantitative agreement with spectra obtained using a conventional FTIR spectrometer. Together with previously reported results, this work demonstrates that QIRS can cover the entire molecular fingerprint region. Our findings highlight the potential of QIRS as a powerful and versatile tool for broadband quantum sensing in the FIR regime. \rev{The present source should also be applicable to other recently demonstrated schemes of QIRS, such as scanless spectral-domain acquisition \cite{Gattinger2026} and open-path QFTIR for remote sensing over long distances \cite{Neves2026}. More generally, the angle-tuned VNIR-FIR photon-pair source is equally applicable to other modalities of photonic quantum sensing, such as quantum imaging \cite{Paterova2020,Kviatkovsky2020,Placke2026} and optical coherence tomography \cite{Paterova2018B,Vanselow2020}, and the broadband, highly nondegenerate phase matching demonstrated here for GaSe is also a key resource for related nonlinear-optical approaches to infrared sensing based on frequency upconversion \cite{Magri2021} and stimulated parametric down-conversion \cite{Dong2025a,Dong2025b}.}

\medskip\noindent\rev{\textit{Note.} After submission of this manuscript, we became aware of a preprint reporting an independent demonstration of spectroscopy with undetected photons up to 20 {\textmu}m using a BaGa$_2$GeSe$_6$ crystal \cite{Dong2026}. Together with the present work, it indicates the growing interest in extending photonic quantum sensing into the long-wavelength infrared region.}

\section*{Funding}

This work was supported by  the Ministry of Education, Culture, Sports, Science and Technology (MEXT) Q-LEAP (Grant No. JPMXS0118067634); Japan Science and Technology Agency, ERATO (Grant No. JPMJER2402), and PRESTO (Grant No. JPMJPR2598); JSPS KAKENHI (Grant No. 24H00195, 21H04444, and 24K06920).

\section*{Acknowledgement}
We would like to thank Y. Taniguchi for his technical support. \rev{Mathematica code used for the calculation of the theoretical photon-pair generation rates (Table 2) was developed with the assistance of Claude (Anthropic; Claude Opus 5.5 and Claude Fable 5.1). All code, numerical results, and physical interpretations were independently validated by the authors. Claude was also used for language editing and organizational suggestions in preparing the manuscript. The tool was not used to generate scientific claims, experimental data, figures, references, or conclusions, and all content was reviewed and verified by the authors.}

\section*{Disclosures}
The authors declare no conflicts of interest.

\section*{Data availability}
Data underlying the results presented in this paper are not publicly available at this time but may
be obtained from the corresponding author (S. Takeuchi) upon reasonable request.

\bibliography{sample_YM_GaSe}

\end{document}